\documentclass[journal,comsoc]{IEEEtran}

\usepackage[T1]{fontenc}
	\usepackage{xcolor}

\ifCLASSINFOpdf
\else
\fi
\usepackage{lipsum}  
\usepackage{amsmath,amsbsy, cite}
\newcommand{\comment}[1]{}
\usepackage[cmintegrals]{newtxmath}
\usepackage{hyperref}

\ifCLASSINFOpdf
\usepackage[pdftex]{graphicx}
\else
\fi

\newcommand\blfootnote[1]{%
  \begingroup
  \renewcommand\thefootnote{}\footnote{#1}%
  \addtocounter{footnote}{-1}%
  \endgroup
}

\begin{document}

\title{On the Performance of UAV Relaying with Reconfigurable Antenna and Media Based Modulation in the Presence of Shadowed Fading}
\author{Ayse~Betul~Buyuksar,~\IEEEmembership{Student~Member,~IEEE,}
Eylem~Erdogan,~\IEEEmembership{Senior~Member,~IEEE,}  
Ibrahim~Altunbas,~\IEEEmembership{Senior~Member,~IEEE}

 }
\maketitle
\begin{abstract}
Unmanned aerial vehicles (UAVs) have attracted significant interest from the academia and industry most recently. Motivated by the wide usage of UAVs, this paper considers UAV communication with reconfigurable antenna (RA) in the presence of fading and shadowing effects which occur due to tall buildings and skyscrapers in urban areas. More precisely, RA offers to receive information through mirror activation patterns (MAPs) so that it can achieve a receive diversity with decreased error probability by using only one radio frequency (RF) chain. {Also, media based modulation (MBM) technique with MAPs can be exploited by using RAs with reduced cost. To quantify the performance of the proposed UAV system, we derive a tight upper bound for the overall error probability by considering approximated channel model based on the standardization studies.} The results have shown that RAs can make the overall system more resilient to shadowing and fading effects in terms of error performance, and they are energy efficient.
\end{abstract}

\IEEEpeerreviewmaketitle

\section{Introduction}

\IEEEPARstart{R}{ecent} advancements have shown that unmanned aerial vehicles (UAVs) will be an important part of next generation non-terrestrial networks as they can provide increased coverage, ultra reliability, instant deployment, and increased throughput with taking part in applications such as cargo delivery, disaster relief, aerial mapping, agricultural irrigation, military attack, and communication relaying \cite{Bithas_2020}. {UAVs can be used as relays on a dual-hop configuration by using amplify-and-forward (AF) and decode-and-forward (DF) techniques to improve the communication reliability and coverage \cite{Zhao_2008}.} UAV relaying is essential when a user equipment ($UE$) and base station ($BS$) are widely scattered or surrounded by densely obstacles including hills and tall buildings. In these densely concentrated areas, UAV relaying can bring significant advantages including enhanced coverage, increased reliability, and high quality of service (QoS) \cite{Zhan_2011}. \blfootnote{A. B. Buyuksar and I. Altunbas are with the Faculty of Electrical and
Electronics Engineering, Istanbul Technical University, 34469 Istanbul, Turkey (e-mail: buyuksar@itu.edu.tr; ibraltunbas@itu.edu.tr).

E. Erdogan is with the Faculty of Engineering and Natural Sciences, Department of Electrical and Electronics Engineering, Istanbul Medeniyet University, 34700 Istanbul, Turkey (e-mail: eylem.erdogan@medeniyet.edu.tr). 

This study was supported by the Scientific and Technological Research Council of Turkey (TUBITAK) under Grant no 117E869.}

One of the most important challenges in UAV relaying is fading and shadowing effects \cite{Khawaja_2019}. To remedy these adverse effects, transmit and/or receive antenna diversity, which can be established by using multiple antennas, can become an important enabler  \cite{Kuhestani_2013}. However, using multiple antennas can increase cost, consumed energy, complexity, and the number of radio frequency (RF) chains. In this regard, reconfigurable antennas (RAs) can be employed to provide receive diversity by using mirror activation patterns (MAPs) without consuming extra energy, and without increasing the number of RF chains. Furthermore, data can be transferred with pattern indices of RAs with only one RF chain \cite{Basar_2017}. {This can decrease the system complexity and allow to use a small and compact communication module for UAV relay systems as described in \cite{Kalis_2008}, and \cite{Ikram_2012}.} In RAs, the transmission of information with the activation or deactivation status of the RF mirrors, named as media-based modulation (MBM) technique, provides better error performance and improved energy/spectral efficiency \cite{Basar_2017}. The MBM concept can be realized by using MAP indices of RAs. In addition to the above-mentioned appealing advantages of MBM systems, significant QoS improvements can be obtained for UAVs \cite{Xu_2019}. Therefore, RAs are expected to satisfy the trade-off between the transmitted power and increased data rate by using MAPs.
 
Using the MAP of RA on a dual-hop UAV relaying system can allow us to achieve receiver diversity at the first hop and enhanced error performance at the second hop with the MBM technique \cite{Can_2021}. In this architecture, receive diversity can be established by selecting the best MAP of RA, which provides the best fading or shadowing channel. The first selection procedure is almost non-applicable for the UAV systems with RA as fading information varies fast, especially in the high-frequency bands \cite{Yilmaz_2013}, \cite{Erdogan_2019}. On the contrary, the latter always varies slowly, which makes it a more practical selection criterion for UAV systems \cite{Erdogan_2021}, \cite{Erdogan_2017}. In the literature, the diversity reception technique based on the concept of shadowing, has been applied in various scenarios, including V2V communications \cite{Bithas_2019} and cooperative systems \cite{Erdogan_2017}. In addition, the diversity reception with beam selection has been proposed in \cite{Maliatsos_2021} for a UAV network.

The current literature is limited with conventional modulation techniques for UAV relays. Specifically, \cite{Chen_2018_1} investigates error performance of different UAV relaying configurations by considering AF and DF relaying under different path loss models over Nakagami-$m$ fading channels. In addition, error performance, outage probabilities, and security concerns are investigated in \cite{Chen_2018_1} and \cite{Letafati_2020} respectively. Despite the energy/spectral efficiency advantages of the MBM technique, none of the works have studied for UAV relays to the best of our knowledge. 

 Another challenge faced by the UAVs is channel modeling \cite{Khawaja_2019}. Standardization studies and the channel measurement campaigns for UAV to ground links showed that multiplicative log-normal shadowing with Rician fading is a suitable model for UAV communications \cite{Khawaja_2019}, \cite{3GPP_Rel15}, and \cite{3GPP_Rel14}. A general distribution that models both fading and shadowing is generalized-$K$ distribution \cite{Shankar}. Furthermore, the log-normal shadowing can be approximated to generalized-$K$ distribution by using the moment matching method \cite{Ahmadi_2010}.

Different from the current literature, this paper proposes a novel UAV relaying model, in which MAPs are used for receive diversity at the first hop and MBM technique at the second hop, on a dual-hop configuration to enhance reliability and performance. More precisely, this paper makes the following contributions:

1) We adopt RA at the UAV relaying system to obtain the receiver diversity based on the shadowing information at the first hop, and so that we can use MBM technique at the second hop. Thereby, the designed model can provide better system performance with reduced complexity for UAV systems. 

2) We derive the error probability for the proposed setup by including path loss, fading and shadowing effects \cite{Khawaja_2019}, \cite{3GPP_Rel14, 3GPP_Rel15} and \cite{ 3GPP_Rel16} with the help of generalized-$K$ channel model which can be adopted for UAV relay systems by applying the approximation methods in \cite[eqn. (2.26)]{Simon_Alouini_2005} and \cite{Peppas_2009}.

3) We provide important guidelines which can be useful for the design of UAV relay systems.

A note on mathematical notations: Vectors and matrices are represented by bold lowercase letters and bold uppercase letters, respectively. $\mathbb{E}[.]$ and $\overline{(.)}$ represent the expectation operator, $f_X(.)$ and $F_X(.)$ denote the probability density function (PDF) and cumulative distribution function (CDF) of $X$, respectively, $|.|$ expresses the absolute value and $\left\| . \right\|_2$ is the Frobenius norm, $\binom{.}{.}$ denotes the binomial coefficient, $\mathcal{CN}(\mu,\sigma^2)$ denotes complex normal distribution with mean $\mu$ and variance $\sigma^2$, $Q\left(.\right)$ denotes the Gaussian-$Q$ function, and $(\boldsymbol{A}\circ\boldsymbol{B})$ expresses the Hadamard products of $\boldsymbol{A}$ and $\boldsymbol{B}$ matrices.	

\section{System and Channel Model}

\subsection{System Model}

\begin{figure}[!htb]
	\centering
	\includegraphics[scale=0.5]{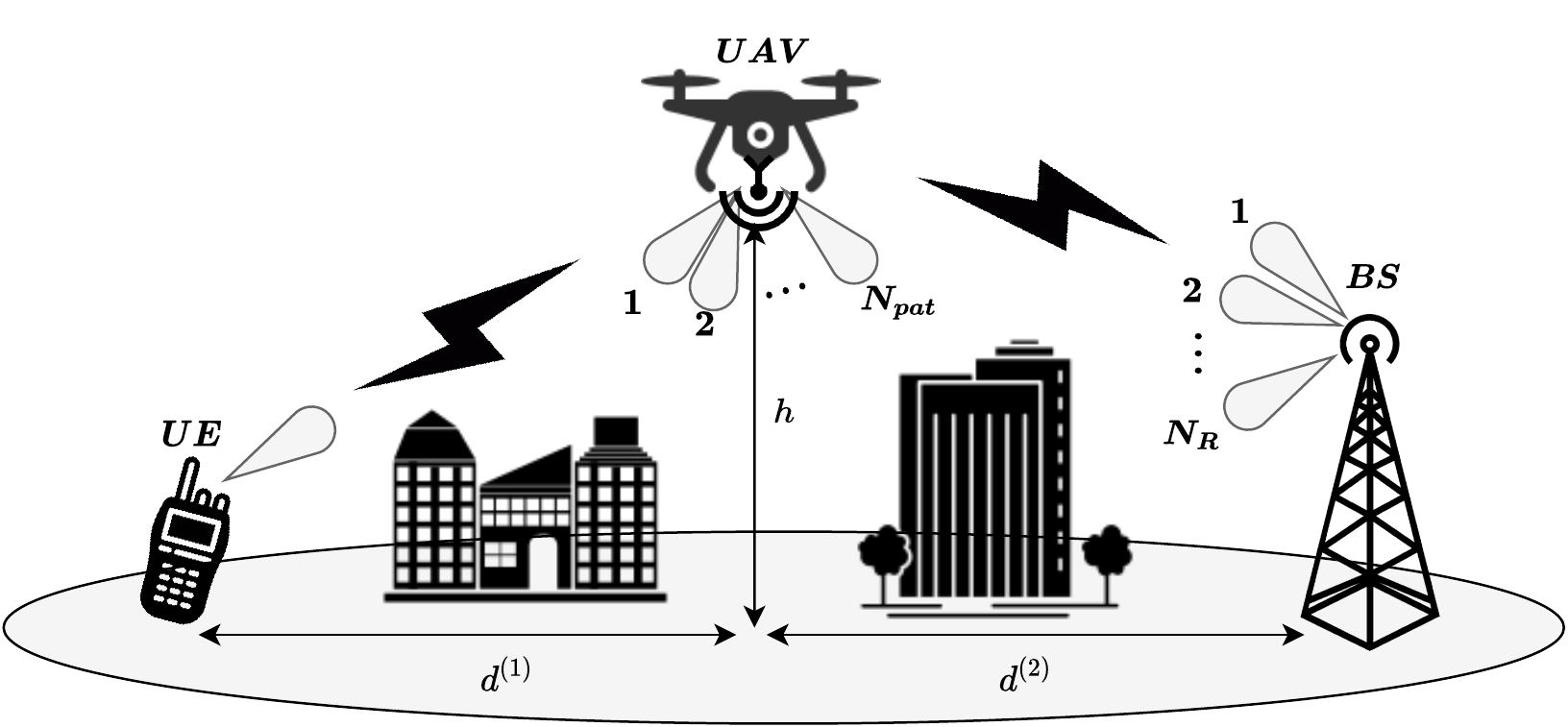}
	\caption{Data transmission from UE to BS through UAV relay with RA.}
	\label{fig:SystModel}
\end{figure}

In this paper, we consider a UAV-aided communication model, as can be seen in Fig. \ref{fig:SystModel}, where a $UE$ with one transmit antenna wishes to communicate with the $BS$ through UAV with RA in the presence of fading and shadowing effects as the $UE$ is located in an urban area that is surrounded with obstacles.\footnote{We assume a quadcopter UAV design which can hover at a specific altitude as widely studied in the literature \cite{Erdogan_2021}, \cite{Wang_2021}.} UAV and $BS$ are equipped with the RA which has $N_{pat}$ MAP and $N_R$ receive antennas, respectively. The direct path between $UE$ and $BS$ cannot be established because of the environmental conditions. Therefore, the communication can be provided with ground to air (GtA) and air to ground (AtG) links. At the first time slot, the received signal at the UAV, \comment{at UAV$r^{(1)}\in \mathbb{C}$}can be expressed as

\begin{equation}
r^{(1)} =\sqrt{\frac{1}{L_{P}^{(1)}}}g_k^{(1)} h_k^{(1)} s^{(1)}+ n^{(1)},
\end{equation}
where $L_{P}^{(1)}$ is the path loss between $UE$ and UAV, $g_k^{(1)}$ is the shadowing coefficient between $UE$ and UAV with the best (say, the $k$th) MAP, which is selected by using shadowing information as \begin{equation}
    g_{k}^{(1)}=\underset{i=1:N_{pat}}{\operatorname{max}} \{g_i^{(1)}\}.
\end{equation}Each shadowing coefficient $g_i^{(1)}$ is modelled as independent identically distributed (i.i.d.) Nakagami-$m_g$ random variable with unit power. The fading coefficients of $UE$ to UAV channel, $h_i^{(1)}$ are given as Nakagami-$m_h$ random variables with uniform phase. Furthermore, $s^{(1)}$ is a $M$-QAM modulated symbol with power ${\mathbb{E}\left[|s^{(1)}|^2\right]=P_{UE}}$, and the  receiver noise at the UAV is modeled as ${n^{(1)} \sim \mathcal{CN}(0,N_0)}$, where $N_0$ is the received noise power. Depending on the  above mentioned information, the signal-to-noise ratio (SNR) with path loss at the first hop can be expressed as ${\gamma=\gamma_{g,\max}\gamma_h}$, where ${\gamma_h=\frac{P_{UE}}{L_P^{(1)}N_0}\left|h_k^{(1)}\right|^2}$ \cite{Erdogan_2019}.\footnote{Standardization and measurement studies in \cite{Khawaja_2019},\cite{3GPP_Rel15}, and \cite{3GPP_Rel14} show that GtA and AtG channels are characterized by shadowed line-of-sight (LOS) propagation with path loss. However, to avoid complicated theoretical analyses \cite{Peppas_2009}, we assume both fading and shadowing coefficients are modeled as i.i.d. Nakagami-$m$ with severity parameter $m_h$ and $m_g$ which means that the channel can be modeled as double Nakagami-$m$ (also called generalized-$K$) distributed \cite{Simon_Alouini_2005},~\cite{Peppas_2009}.} After the first transmission period, UAV decodes the $M$-QAM signal by using maximum-likelihood (ML) detector which can be given as
\begin{equation}
\hat{s}_{\ell}^{(1)}=\underset{i=1:M}{\operatorname{argmin}}\left|r^{(1)}- \sqrt{\frac{1}{L_{P}^{(1)}}}g_k^{(1)} h_k^{(1)} \tilde{s}_i^{(1)} \right|^2,
\end{equation} where $\tilde{s}_i^{(1)}$ is the $i$th $M$-QAM symbol. Due to complexity considerations at UAV, we assume that UAV relay employed with the RA antenna with one RF chain. Therefore, decoded data at UAV matches with the corresponding pattern indice of the RA which can be shown as $M=N_{pat}$. At the second time slot, the received signal at the $BS$ can be expressed as 
\begin{equation}
\boldsymbol{r}^{(2)} =\sqrt{\frac{1}{L_{P}^{(2)}}}\left(\boldsymbol{G}^{(2)} \circ\boldsymbol{H}^{(2)}\right) \boldsymbol{s}_{\ell}^{(2)}+ \boldsymbol{n}^{(2)},
\end{equation}
where $\boldsymbol{s}_{\ell}^{(2)}\in \mathbb{R}^{N_{pat}\times 1}$ is the transmission vector whose only the $\ell$th element is non-zero with power ${\mathbb{E}\{||\boldsymbol{s}_{\ell}^{(2)}||_2^2\}=P_{U\!AV}}$ which means only the $\ell$th MAP in transmission. Moreover, $L_{P}^{(2)}$ is the path loss between UAV to $BS$, ${\boldsymbol{G}^{(2)}\in \mathbb{R}^{N_R\times N_{pat}}}$ is the shadowing coefficients matrix between UAV and $BS$ whose elements are modelled as i.i.d. Nakagami-$m_g$ distributed random variables, ${\boldsymbol{H}^{(2)}\in \mathbb{C}^{N_{R}\times N_{pat}}}$ is the fading coefficients matrix, which is generated as i.i.d. Nakagami-$m_h$ with uniform phase, and additive noise vector $\boldsymbol{n}^{(2)} \in \mathbb{C}^{N_R \times 1}$ has components which are distributed as ${ \mathcal{CN}(0,N_0)}$ where $N_0$ is the received noise power at $BS$. Finally, $BS$ decodes the signal by using ML  decoding rule as
\begin{equation}
\hat{s}_{\ell}^{(2)}=\underset{i=1:N_{pat}}{\operatorname{argmin}}\left\|\boldsymbol{r}^{(2)} -\sqrt{\frac{1}{L_{P}^{(2)}}}\left(\boldsymbol{G}^{(2)} \circ\boldsymbol{H}^{(2)}\right) \tilde{\boldsymbol{s}}_i^{(2)} \right\|_2^2,
\end{equation}
where $\tilde{\boldsymbol{s}}_i^{(2)}$ is a vector whose only the $i$th element is non-zero.

In the considered dual-hop system, two time slots are required to transmit $\log_2M$ bits. Therefore, the bandwidth efficiency of the system is $\eta=\left(\log_2M\right)/2$ bit/s/Hz. Since we assumed that $M=N_{pat}$, the bandwidth efficiency of the system increases logarithmically with the number of MAPs at the UAV's RA. With the increase of $N_{pat}$, the bandwidth efficiency of the overall system increases, and receiving antenna diversity gain is obtained as a result of pattern selection at the first hop. Therefore, both bandwidth efficiency and error performance of the system can be improved by means of $N_{pat}$ depending on the system and channel parameters.

\subsection{The UAV Relaying Path Loss Model}

The path loss model for the GtA and AtG communication ($L_P^{(i)}$, $i\in\{1,2\}$) can be given very similar to \cite{Chen_2018_1} and \cite{Hourani_2014} as
\begin{equation}
\begin{split}
L_P^{(i)}=&\beta^{(i)}\left(\sqrt{h^2+(d^{(i)})^2}\right)^{\alpha},
\end{split}
\label{eq:PathLoss}
\end{equation}
where $\alpha$, $h$, $d^{(i)}$ are path loss exponent, height of the UAV and vertical distance of air and ground unit, respectively and $ \beta^{(i)}=10^{\frac{B}{10}+\frac{A}{10+10a'e^{-b'(\theta^{(i)}-a')}}}$. Here, $A$ and $B$ are defined as ${A=\eta_{LOS}-\eta_{NLOS}}$ and ${B=20\log_{10}\left(\frac{4\pi f_c}{c}\right)+\eta_{NLOS}}$. $f_c$ is the carrier frequency and $c$ is the speed of light and the angle $\theta^{(i)}$ can be calculated as ${\theta^{(i)}=\arctan\left(h/d^{(i)}\right)}$. Furthermore, $a'$, $b'$, $\eta_{LOS}$, $\eta_{NLOS}$ parameters are determined in \cite{Hourani_2014} for different environments as can be seen in Table \ref{table:environment}.
{\renewcommand{\arraystretch}{1.3}
\begin{table}[!tb]
	\begin{center}
		\caption{Path loss parameters for different environments.}
		\label{tab:table1}
		\begin{tabular}{|l|c|c|c|c|}
			\hline
			Environment &$\eta_{LOS}$ & $\eta_{NLOS}$ & $a'$&$b'$ \\
			\hline
			Suburban 	& 0.1 	& 21	& 4.88    & 0.429 \\
			Urban 		& 1 	& 20	& 9.6117  & 0.1581 \\
			Dense Urban & 1.6 	& 23	& 12.081  & 0.1139 \\
			Highrise Urban &2.3 & 34 	& 27.2304 & 0.0797 \\
			\hline
		\end{tabular}
		\label{table:environment}
	\end{center}
	\vspace{-0.5cm}
\end{table}}

\section{PERFORMANCE ANALYSIS}

In this section, we analyze the overall symbol error probability (SEP) of the proposed system. The end-to-end (e2e) error probability for dual-hop regenerative systems is dominated by the weakest hop \cite{Hasna_2004} and therefore we define the SEP as
\begin{equation}
\begin{split}
P_{s}(\epsilon)\approx \max \{P_{s,1}(\epsilon),P_{s,2}(\epsilon)\}.
\end{split}
\label{eq:Ps}
\end{equation}
Herein, $P_{s,1}(\epsilon)$ and $P_{s,2}(\epsilon)$ are SEP of  the GtA and the AtG links, respectively.\footnote{To obtain a closed form SEP expression and to verify the result, we have used integer-valued fading severity parameters.} 

\subsection{GtA Communication}
 The SEP for the GtA link can be written as \cite{Erdogan_2017}
 \begin{equation}
\begin{split}
P_{s,1}(\epsilon)=&\frac{C}{2}-\frac{C\sqrt{D}}{2\sqrt{\pi}}\int_{0}^{\infty}\frac{\exp(-D\gamma)}{\sqrt{\gamma}}\overline{F}_{\gamma}(\gamma)d\gamma,
\label{eq:SEP_OverCDF}
\end{split}
\end{equation} 
 where $C=4\left(1-\frac{1}{\sqrt{M}}\right)$, $D=\frac{3}{2(M-1)}$ for the Gray-mapped $M$-QAM and $C=1$, $D=1$ for BPSK modulation schemes. $\overline{F}_{\gamma}(\gamma)=1-F_{\gamma}(\gamma)$ is the complementary CDF and $F_{\gamma}(\gamma)$ can be given as \cite{Yilmaz_2013}, \cite{Erdogan_2019} 
\begin{equation}
\begin{split}
F_{\gamma}(\gamma)=\int_{0}^{\infty}F_{\gamma_h}(\gamma/u)f_{\gamma_{g,\max}}(u)du.
\end{split}
\label{eq:1stCDF}
\end{equation}
 
\textbf{Lemma 1.} By substituting  \eqref{eq:1stCDF} into \eqref{eq:SEP_OverCDF}, $P_{s,1}(\epsilon)$ can be obtained for integer values of $m_g$ and $m_h$ as shown in \eqref{eq:TAS_SEP} at the top of the next page. Note that, $\Gamma(.)$ and $W_{.,.}( . )$ denote the Gamma and the Whittaker function, respectively \cite[eqn. (8.310) \text{ and } (9.222)]{gradshteyn2007}.

\textbf{Proof} See Appendix \ref{appendix:a}.

\begin{figure*}
	\begin{equation}
	\begin{split}
	P_{s,1}(\epsilon)
	=&\frac{C}{2}-\frac{C\sqrt{D}}{2\sqrt{\pi}}\frac{N_{pat}}{\Gamma(m_g)}\left(\frac{m_g}{\overline{\gamma}_g}\right)^{m_g}\sum_{r=0}^{N_{pat}-1}\sum_{p=0}^{r(m_g-1)}\sum_{s=0}^{m_h-1}\binom{N_{pat}-1}{r}\frac{m_h^s}{s!} \chi_p^r(-1)^r \left(\frac{m_h{\Omega_1}}{m_g(r+1)}\right)^{\frac{m_g-s+p}{2}} \\
	&\frac{\Gamma\left(m_g+p+1/2\right)\Gamma\left(s+1/2\right)}{2\sqrt{\frac{m_gm_h(r+1)}{2D{\Omega_1}}}}\exp\left(\frac{m_gm_h(r+1)}{2D{\Omega_1}}\right)D^{-\frac{m_g-s+p}{2}}W_{-\frac{m_g+s+p}{2}, \ \frac{m_g-s+p}{2}}\left(\frac{m_gm_h(r+1)}{D{\Omega_1}}\right).
	\end{split}
	\label{eq:TAS_SEP}
	\end{equation} 
	\hrulefill
	\vspace{-0.2cm}
\end{figure*}

\subsection{AtG Communication}

As we apply MBM at the UAV, the SEP for the AtG link can be tightly upper bounded as \cite{Peppas_2014}
\begin{equation}
\begin{split}
P_{s,2}(\epsilon) \leq&\frac{N_{pat}\log_2(N_{pat})}{2}\mathbb{E}\left[\Pr\left(\boldsymbol{s}_{\ell}^{(2)}\!\! \rightarrow \boldsymbol{s}_{\upsilon}^{(2)}\bigg|\left(\boldsymbol{G}^{(2)} \! \circ \! \boldsymbol{H}^{(2)}\right)\right)\right],
\end{split}
\label{eq:2ndSEPNumeric}
\end{equation}
 where  $\Pr\left(\boldsymbol{s}_{\ell}^{(2)}\rightarrow \boldsymbol{s}_{\upsilon}^{(2)}\bigg|\left(\boldsymbol{G}^{(2)} \circ\boldsymbol{H}^{(2)}\right) \right)$ expresses the conditional pairwise error probability (CPEP) for the erroneous detection of $\boldsymbol{s}_{\ell}^{(2)}$ to  $\boldsymbol{s}_{\upsilon}^{(2)}$ with $\ell\neq\upsilon$. Here CPEP can be written~as
\begin{equation}
\begin{split}
\Pr&\left(\boldsymbol{s}_{\ell}^{(2)}\rightarrow \boldsymbol{s}_{\upsilon}^{(2)}\bigg|\left(\boldsymbol{G}^{(2)} \circ\boldsymbol{H}^{(2)}\right)\right) \\
=& \ Q\left(\sqrt{\frac{P_{U\!AV}}{2L_P^{(2)}N_0} \left\| \left(\boldsymbol{g}_{\ell}^{(2)} \circ\boldsymbol{h}_{\ell}^{(2)}\right)-  \left(\boldsymbol{g}_{\upsilon}^{(2)} \circ\boldsymbol{h}_{\upsilon}^{(2)}\right)\right\|_2^2}\right),
\label{eq:2ndSEPNumeric_1}
\end{split}
\end{equation}
where i.i.d. $\boldsymbol{g}_{\ell}^{(2)}$ and $\boldsymbol{h}_{\ell}^{(2)}$ vectors express the $\ell$th column of the $\boldsymbol{G}^{(2)}$ and $\boldsymbol{H}^{(2)}$ matrices. Since all channel coefficients are i.i.d. random variables, the SEP for the AtG link can be expressed by using Craig's formula \cite[eqn. (4.2)]{Simon_Alouini_2005},
\begin{equation}
P_{s,2}(\epsilon)\!\! \leq \frac{N_{pat}\log_2(N_{pat})}{2\pi}\!\int_0^{\pi/2}\!\!\!\!\!\! M_{\Phi}\left(\frac{{\Omega_2}}{4\sin^2(\varphi)} \right)^{N_R}\!\!\!\!d\varphi,
\label{eq:2ndSEPNumeric_2}
\end{equation}
where $\Omega_2=\frac{P_{U\!AV}}{L_P^{(2)}N_0}$, $M_{\Phi}\left(.\right)$ is the moment generating function of ${\Phi =\left|g_{1,\ell}^{(2)}  h_{1,\ell}^{(2)}-  g_{1,\upsilon}^{(2)}  h_{1,\upsilon}^{(2)}\right|^2} $. Here, $g_{1,\ell}^{(2)} $ and $ h_{1,\ell}^{(2)}$ express the elements of $\boldsymbol{G}^{(2)}$ and $\boldsymbol{H}^{(2)}$ matrices at the first row of the $\ell$th column. $\Phi$ does not depend on $\ell$ and $\upsilon$ since all $g_{1,\ell}$, $ h_{1,\ell}$, $g_{1,\upsilon}$ and $h_{1,\upsilon}$ coefficients are i.i.d. random variables. To the best of Authors' knowledge, \eqref{eq:2ndSEPNumeric_2} cannot be expressed in closed-form. By using the asymptotical approach that is used in \cite{Peppas_2014}, asymptotic SEP expression for the AtG link becomes
\begin{equation}
\begin{split}
P_{s,2}(\epsilon) \leq&\frac{N_{pat}\log_2(N_{pat})2^{N_R-2}\left(2N_R\right)!}{(N_R!)^2} \left(\frac{\Omega_2}{\zeta}\right)^{-N_R}
\end{split}
\label{eq:2ndSEPAsymp}
\end{equation}
where 
${
\zeta=\frac{m_gm_h}{\left(\Gamma (m_g)\right)^2\left(\Gamma (m_h)\right)^2}G_{2,2}^{2,2} \left(
\begin{array}{c}  1  \end{array}
\middle\vert
\begin{array}{c}
1-m_h,1-m_g\\
m_h-1,m_g-1\\
\end{array}
\right)
}$,
and $G_{c,d}^{a,b} \left(
\begin{array}{c}  .  \end{array}
\middle\vert
\begin{array}{c}
.,.\\
.,.\\
\end{array}
\right)$ denotes the Meijer-G function \cite[eqn. (9.307)]{gradshteyn2007}. By substituting \eqref{eq:2ndSEPAsymp} and \eqref{eq:TAS_SEP} into \eqref{eq:Ps} $P_s(\epsilon)$ can be obtained for the proposed system.

\subsection{Diversity Order Analysis}

To obtain the diversity order of the system, we have used the high-SNR approximation technique presented in \cite{Wang_2003}. At high SNR, the asymptotic SEP can be expressed as 
\begin{equation}
    {P_s^{\infty}\approx \max \{\left(G_{a,1}\Omega_1\right)^{G_{d,1}},\left(G_{a,2}\Omega_2\right)^{G_{d,2}}\}},
\end{equation}
where $G_{a,1}$, $G_{d,1}$, $G_{a,2}$, $G_{d,2}$ are the array and diversity gains of the first and second hops, respectively. The array and diversity gains of the first hop can be obtained by using the asymptotic CDF expression as  
$ F_{\gamma}^{\infty}(\Omega_1)=\Upsilon \left(\Omega_1\right)^{-G_{d,1}}$ \cite[eqn. (45:9:1)]{Oldham_2008}. Here,
\begin{equation}
\Upsilon = \left\{
    \begin{array}{ll}
        \frac{N_{pat}\Gamma{(N_{pat}m_g-m_h)}(m_gm_h)^{m_h}}{m_g^{N_{pat}-1}\Gamma(m_g)^{N_{pat}}\Gamma(m_h+1)}, & N_{pat}m_g>m_h \\
         \frac{N_{pat}(m_gm_h)^{\frac{N_{pat}m_g+m_h}{2}}\log\left(\frac{\psi}{m_gm_h}\right)}{m_g^{N_{pat}-1}\Gamma(m_g)^{N_{pat}}\Gamma(m_h)},& N_{pat}m_g=m_h\\
       \frac{\Gamma{(m_h-N_{pat}m_g)}(m_gm_h)^{N_{pat}m_g}}{m_g^{N_{pat}}\Gamma(m_g)^{N_{pat}}\Gamma(m_h)} ,& N_{pat}m_g<m_h
    \end{array}
\right. ,
\end{equation}
where $\psi$ denotes Euler's constant. After some manipulations and by using $F_{\gamma}^{\infty}(\Omega_1)$, asymptotic SEP $P_{s,1}^{\infty}(\epsilon)$ can be expressed as  $  P_{s,1}^{\infty}(\epsilon)=\frac{C\Upsilon\Gamma\left(G_{d,1}+1/2\right)}{2\sqrt{\pi}\left(D\Omega_1\right)^{G_{d,1}}}$ \cite{Yilmaz_2013}, \cite{Erdogan_2017}.
 The resulting array and diversity gains of the first hop can be written as  \begin{equation}
\begin{split}
         G_{a,1}&=  \left(\frac{C\Upsilon\Gamma{\left(G_{d,1}+1/2\right)}}{2D\sqrt{\pi}}\right)^{G_{d,1}},\\
         G_{d,1}&=\min\{N_{pat}m_g,m_h\}.
\end{split}
\end{equation}
The array gain and the diversity gain of the second hop can be easily obtained by using \eqref{eq:2ndSEPAsymp} as
\begin{equation}
\begin{split}
         G_{a,2}&=  \frac{N_{pat}\log_2(N_{pat})2^{N_R-2}\left(2N_R\right)!}{(N_R!)^2} \left(\zeta\right)^{N_R},\\
         G_{d,2}&=N_R.
\end{split}
\end{equation}
 Thus, the overall diversity order of the system can be written as ${G_d = \min\{\min\{N_{pat}m_g, m_h\}, N_R\}}$. 

\section{Numerical results}

In this section, we first verify the theoretical findings by a set of Monte-Carlo simulations. Thereafter, the error performance of the proposed scheme is presented for different modulation levels, fading and shadowing severity parameters. Finally, we provide some important guidelines that can be used in the design of UAV systems. 

In the simulations, we set $UE$ and UAV signal powers as ${P_{UE}=P_{U\!AV}=23}$ dBm \cite[Table A.1-1]{3GPP_Rel15}, \cite[Table 5.3.3-1]{3GPP_Rel16} and therefore total transmitted power becomes ${P_T=P_{UE}+P_{U\!AV}=26}$ dBm. The carrier frequency, the white noise power density and bandwidth of the communication system are set to ${f_c=2}$ GHz, ${-174}$ dBm/Hz, ${BW=10}$ MHz respectively as in \cite[Table 5.3.3-1]{3GPP_Rel16}. Furthermore, the UAV is deployed at equal distance to the $UE$ and BS ($d^{(1)}=d^{(2)}$) with $h=100$ m, therefore $L_P={L_P^{(1)}=L_P^{(2)}}$.

By using logaritmic standard variation of the log-normal shadowing effect, ${\sigma_{dB}=4\text{ dB}, \ 6\text{ dB}, \ 8\text{ dB}}$ in \cite{3GPP_Rel14}, the parameter of the Nakagami-$m_g$ distribution, which is approximated with moment matching method, can be found as $m_g\approx 4, \ 2,\ 1$ \cite{Peppas_2009}. The fading effect in the GtA/AtG links given as Rician fading in \cite{Khawaja_2019} and \cite{3GPP_Rel14} with $K=5$ dB and $K=10$ dB, can be easily and closely approximated as Nakagami-$m_h$ distribution with $m_h\approx 2$ and $6$ \cite{Bithas_2020}, \cite[eqn.~(2.26)]{Simon_Alouini_2005}.

\subsection{Verifications of the Theoretical Expressions}

Fig. 2 verifies the theoretical findings with simulations, where SEP is depicted as a function of total transmitted power ($P_T$) over path loss ($L_P$) and noise power ($N_0$) per link. The performance is presented for heavy shadowing ($m_g=1$) and fading with $m_h=6$. As we can see, the theoretical upper bound for SEP curves which are shown with solid lines are in good agreement with the marker symbols which are generated by the simulations for different modulation types. Moreover, the asymptotic curves are close to the theoretical ones for the system using $BPSK$ and $4-QAM$ modulations. However, the asymptotic curve  of the $16-QAM$ system is a bit far from the theoretical curve due to increased $M$ and $N_{pat}$ values. One can easily see that the overall diversity gains and bandwidth efficiencies are $2,\ 4,\ 6 $ and $1/2,\ 1,\ 2$ bit/s/Hz for $BPSK$, $4-QAM$, and $16-QAM$ systems, respectively. Furthermore, the $16-QAM$ system has the best performance at high SNR while providing the best bandwidth efficiency. In the proposed system, the received signal at the UAV's RA is obtained from the best MAP which has the maximum shadowing information. Therefore, increasing $N_{pat}$ leads to improved error performance as we assume $M=N_{pat}$.
\begin{figure}[!htb]
	\centering
	\includegraphics[scale=0.4]{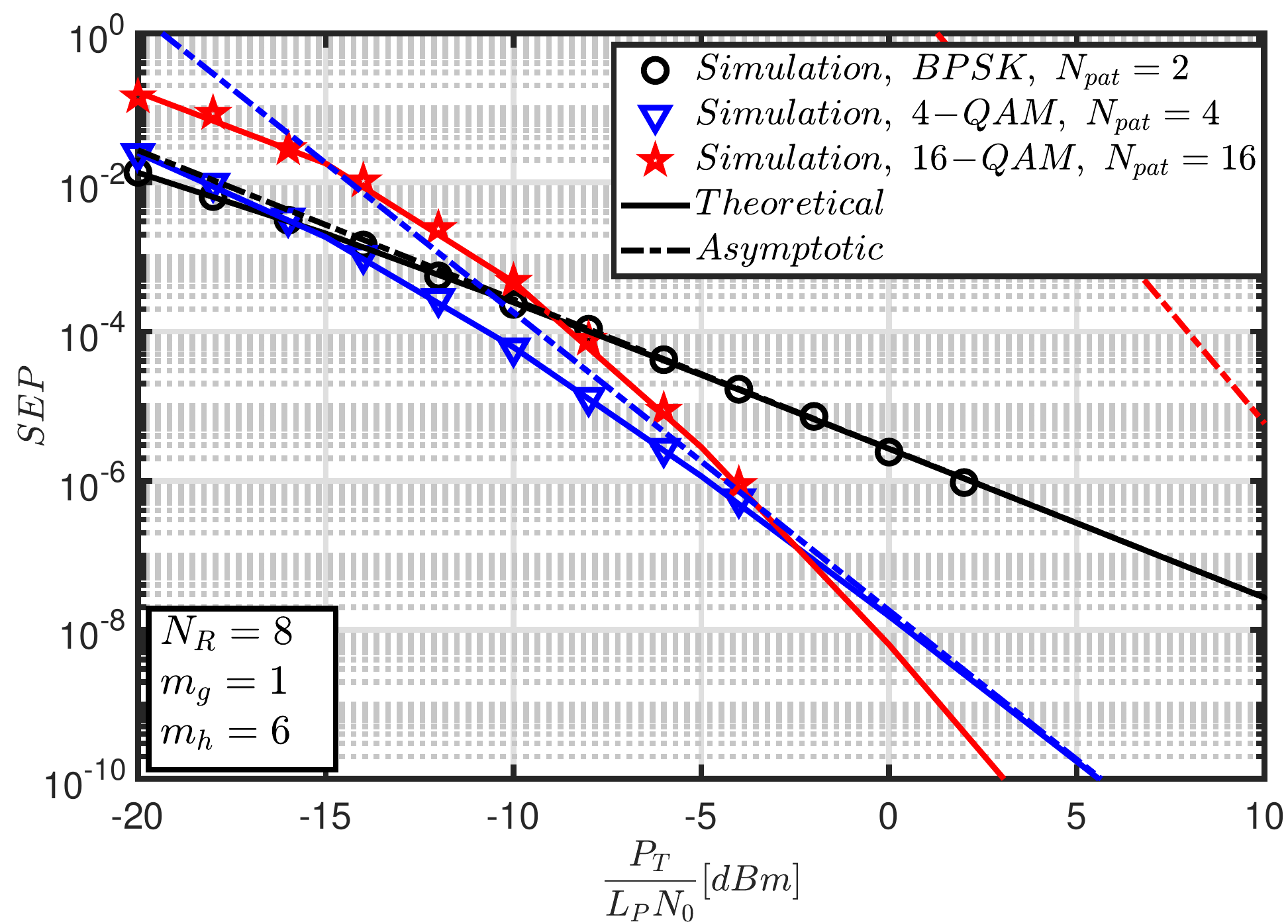}
	\caption{SEP analysis of the system with heavy shadowing and fading with $m_h=6$.}
	\label{fig:1}
\end{figure}
\subsection{Comparison of Error Performances under Different Environments}

\begin{figure}[!tb]
	\centering
	\includegraphics[scale=0.4]{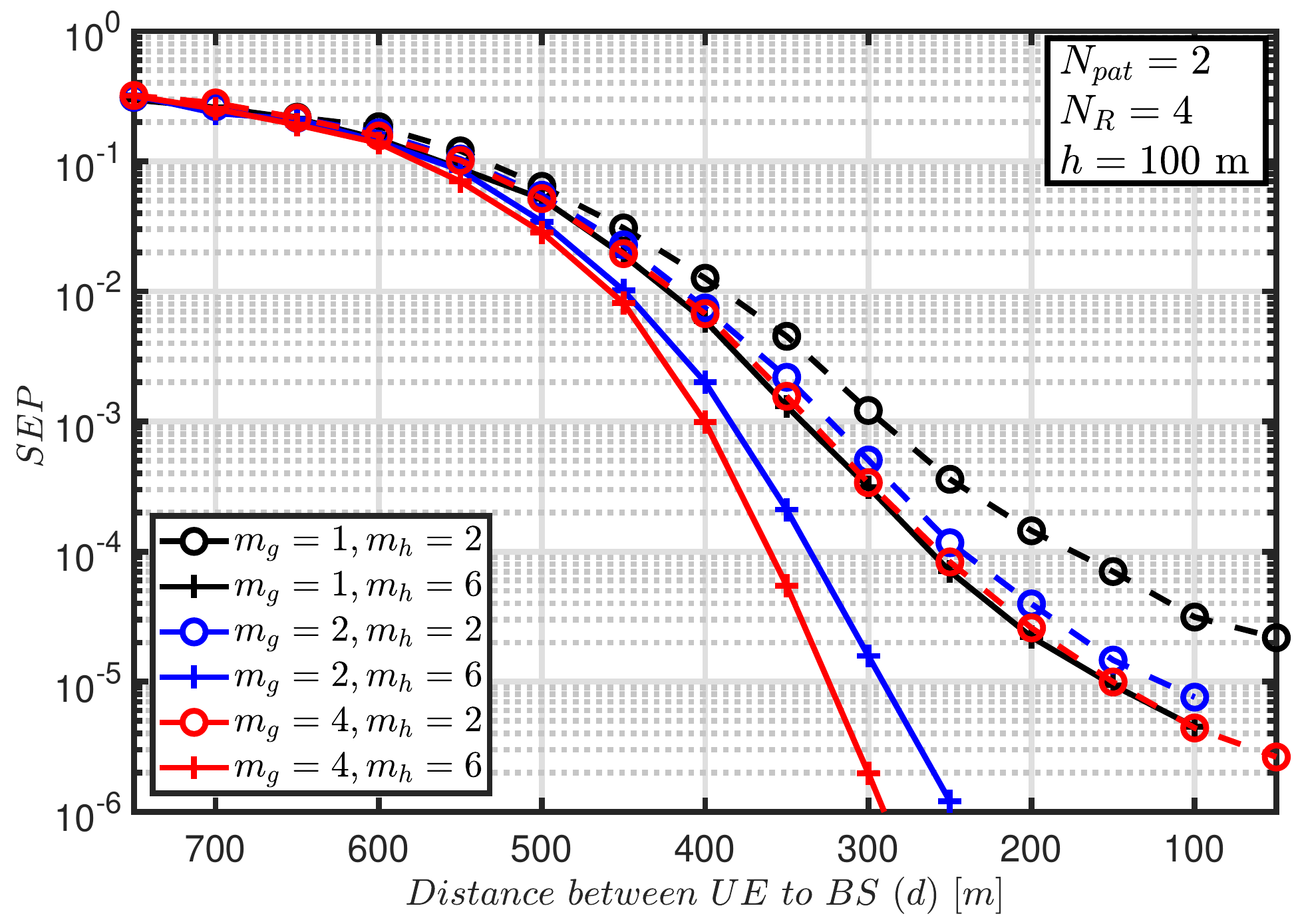}
	\caption{SEP analysis of the system in urban area for the different shadowing and fading effects when BPSK symbols transmitted from $UE$.}
	\label{fig:2}
\end{figure}

Fig. \ref{fig:2} depicts the SEP for BPSK symbols as a function of the distance $d=d^{(1)}+d^{(2)}$ between $UE$ and $BS$ by considering Table \ref{tab:table1}. As can be observed from the figure, in the presence of heavy shadowing ($m_g=1$), the error performance of the proposed scheme degrades. The figure also shows that fading affects the error performance more with respect to the~shadowing.  

\begin{figure}[!tb]
	\centering
	\includegraphics[scale=0.38]{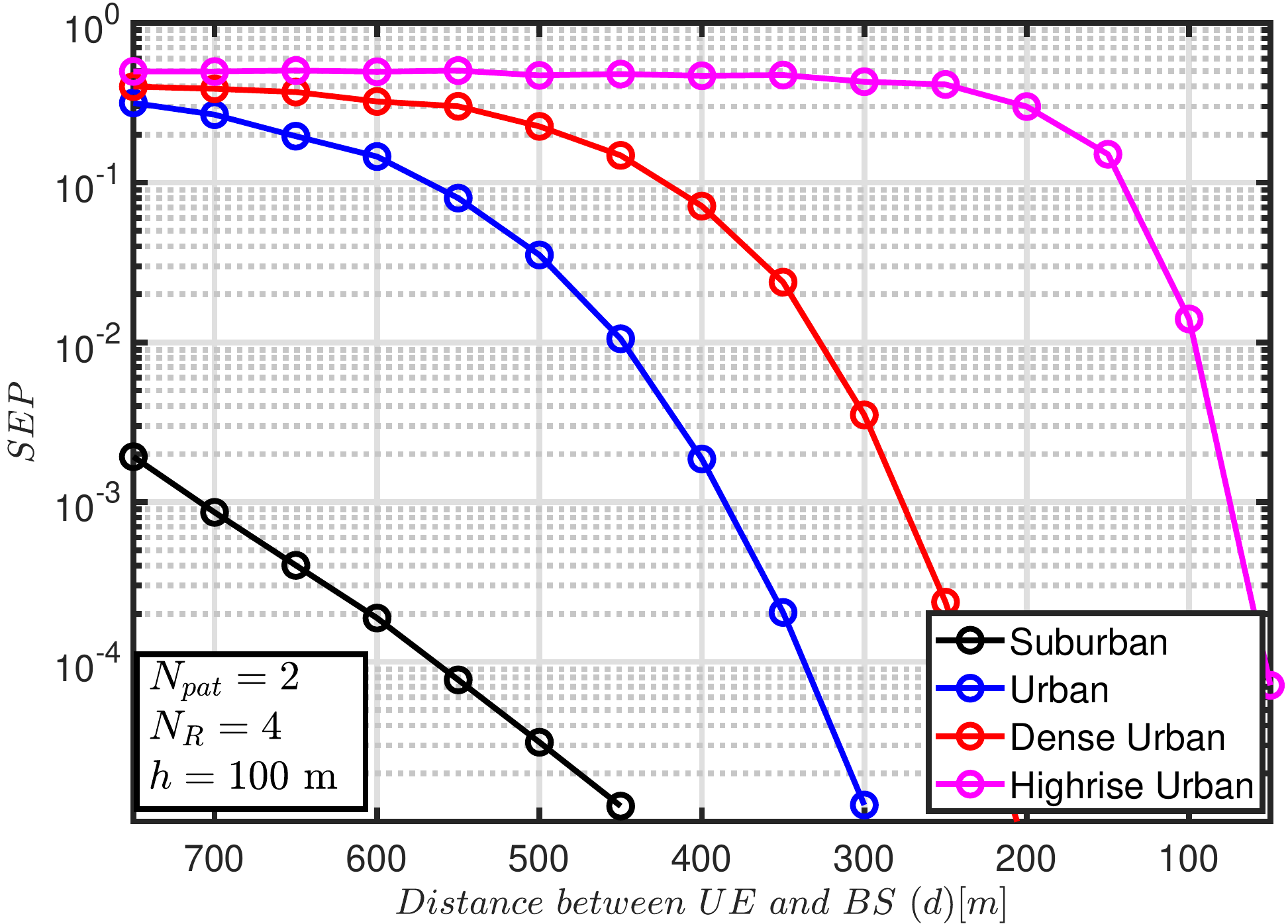}
	\caption{SEP analysis of the system in different environments when BPSK symbols transmitted from $UE$, strong shadowing effect $m_g=1$ and fading with $m_h=2$.}
	\label{fig:3}
\end{figure}

Fig. \ref{fig:3} shows the SEP curves of BPSK symbols as a  function of distance between $UE$ and $BS$ ($d$) for different environments by considering Table \ref{tab:table1}. As expected, the system performance enhances in suburban environment and diminishes in highrise urban environment.

\subsection{Design Guidelines}

In this subsection, we shed light on the design of practical UAV relaying with RA system by providing some important design guidelines.
\begin{itemize}
\item UAV relaying with RA system can be preferable as it can adaptively adjust the data transmission rate and the error performance without increasing the cost of RF chain and the number of antennas.

\item Using shadowing information to select the MAP of RAs can improve spectral efficiency by eliminating fast fading channel estimation symbols. Furthermore, shadowing information-based MAP selection is more practical and efficient for UAVs as shadowing information varies~slowly.

\item Since e2e SEP performance of the UAV relay system is determined by the weakest hop, it is important to improve communication link reliability by considering the environmental conditions, channel shadowing and fading effects to determine the weakest hop. Thereby, RAs provides flexibility to system by increasing error performance of the weakest hop. 
\end{itemize}

\section{Conclusion}
As the MBM technique with reconfigurable antennas supports cost and energy-efficient transmission with high bandwidth efficiency, we have proposed a novel UAV relaying system with MBM and analyzed its error performance by using power limitations and channel model parameters which are determined in standardization studies. To quantify the performance of the proposed scheme, we have derived a tight upper bound for error performance expression. The results show that reconfigurable antennas can bring many advantages to the UAV relaying systems as they can provide increased system reliability, and bandwidth efficiency. Moreover, considered system can be investigated for correlated RA channels, space-time codes with multiple RA, and security aspects as future work.

\appendices
\section{}
\label{appendix:a}
 For integer-valued $m_h$, the CDF expression of Gamma distribution can be written by using \cite[eqn. (8.352.6)]{gradshteyn2007} as
\begin{equation}
\begin{split}
F_{\gamma_h}(\gamma/u)=&1-\exp\left(-\frac{\gamma m_h}{u{\Omega_1}}\right)\sum_{p=0}^{m_h-1}\frac{1}{p!}\left(\frac{\gamma m_h}{u{\Omega_1}}\right)^p,
\end{split}
\label{eq:Gamma_h_CDF}
\end{equation} 
where ${\Omega_1}=\frac{P_{UE}}{L_P^{(1)}N_0}$ and the PDF of the $\gamma_{g,\max}=\left|g_k^{(1)}\right|^2$ can be expressed as
${f_{\gamma_{g,\max}}(u)=N_{pat}(F_{\gamma_g}(u))^{N_{pat}-1}f_{\gamma_g}(u)}$,
where PDF of the Gamma-distributed $\gamma_{g}=\left|g_i^{(1)}\right|^2$ and can be written~as
\begin{equation}
\begin{split}
f_{\gamma_g}(u)=&\left(\frac{m_g}{\overline{\gamma}_g}\right)^{m_g}\frac{1}{\Gamma(m_g)}x^{m_g-1}\exp\left(-\frac{um_g}{\overline{\gamma}_g}\right)
\end{split}
\end{equation}
where $\overline{\gamma}_g=\mathbb{E}[|g_i^{(1)}|^2]=1$. Therefore by considering integer-valued $m_g$ and with the aid of binomial expansion \cite[eqn. (0.314)]{gradshteyn2007}, we can write $\left(F_{\gamma_g}(u)\right)^{N_{pat}-1}$ as
\begin{equation}
\small
\begin{split}
\left(F_{\gamma_g}(u)\right)^{N_{pat}-1}\!\!=& \!\!\! \sum_{r=0}^{N_{pat}-1}\!\!\binom{N_{pat}-1}{r}(-1)^r\exp\left(-\frac{urm_g}{\overline{\gamma}_g}\right)\!\!\sum_{p=0}^{r(m_g-1)}\!\!\!\chi_p^r u^p.
\end{split}
\label{eq:1stCDF2}
\end{equation}
  Here $\chi_p^r$ stands for the multinomial coefficients and given as $\chi_p^r=\frac{1}{pa_0}\sum_{n=1}^{p}(nr-p+n)a_n\chi_{p-n}^r$, $\chi_0^r=a_0^r$ where ${a_n=\frac{1}{n!}\left(\frac{m_g}{\overline{\gamma}_g}\right)^n}$. By using \eqref{eq:1stCDF2}, the PDF of $f_{\gamma_{g,\max}}(u)$ can be expressed as
\begin{equation}
\begin{split}
f_{\gamma_{g,\max}}(u)=&\left(\frac{m_g}{\overline{\gamma}_g}\right)^{m_g}\frac{N_{pat}}{\Gamma(m_g)}\sum_{r=0}^{N_{pat}-1}\sum_{p=0}^{r(m_g-1)}\binom{N_{pat}-1}{r}\\
&(-1)^r\chi_p^r\exp\left(-\frac{um_g(r+1)}{\overline{\gamma}_g}\right) u^{m_g+p-1}.
\end{split}
\label{eq:1stCDF3}
\end{equation}
By using \eqref{eq:Gamma_h_CDF} and \eqref{eq:1stCDF3} in \eqref{eq:1stCDF}, and with the aid of \cite[eqn. (3.471.9)]{gradshteyn2007}, $F_{\gamma}(\gamma)$ can be written as
\begin{equation}\small
\begin{split}
F&_{\gamma}(\gamma)
=1-\frac{N_{pat}}{\Gamma(m_g)}\left(\frac{m_g}{\overline{\gamma}_g}\right)^{m_g}\sum_{r=0}^{N_{pat}-1}\sum_{p=0}^{r(m_g-1)}\sum_{s=0}^{m_h-1}\binom{N_{pat}-1}{r}\frac{m_h^s}{s!}\\ 
& \chi_p^r(-1)^r 2\left(\frac{\gamma m_h\overline{\gamma}_g}{m_g(r+1){\Omega_1}}\right)^{\frac{m_g-s+p}{2}}K_{m_g+p-s}\left(2\sqrt{\frac{\gamma m_hm_g(r+1)}{\Omega_1\overline{\gamma}_g}}\right)
\end{split}
\label{eq:TAS4}
\end{equation}
where $K_{.}(.)$ denotes modified Bessel function of the second kind \cite[eqn. (8.407)]{gradshteyn2007}. The SEP expression for the GtA link can be found by substituting \eqref{eq:TAS4} in \eqref{eq:SEP_OverCDF}.
\ifCLASSOPTIONcaptionsoff
  \newpage
\fi

\end{document}